\documentstyle[version2,preprint,aps]{revtex}
\def\ut#1{\rlap{\lower1ex\hbox{$\sim$}}#1{}}
\begin{document}
\begin{title}
{\bf JONES POLYNOMIALS FOR INTERSECTING KNOTS \\ AS PHYSICAL
STATES OF QUANTUM GRAVITY}
\end{title}
\author{Bernd Br\"ugmann}
\begin{instit}
Physics Department, Syracuse University,
Syracuse, NY 13244
\end{instit}
\author{Rodolfo Gambini}
\begin{instit}
Instituto de F\'{\i}sica, Facultad de Ingenier\'{\i}a,\\
J. Herrera y Reissig 565, C.C. 30, Montevideo, Uruguay.
\end{instit}
\moreauthors{Jorge Pullin}
\begin{instit}
Department of Physics, University of Utah,
Salt Lake City, UT 84112
\end{instit}
\begin{abstract}

We find a consistent formulation of the constraints of Quantum Gravity
with a cosmological constant in terms of the Ashtekar new variables in
the connection representation, including the existence of a state that
is a solution to all the constraints.  This state is related to the
Chern-Simons form constructed from the Ashtekar connection and has an
associated metric in spacetime that is everywhere nondegenerate.  We
then transform this state to the loop representation and find
solutions to all the constraint equations for intersecting loops.
These states are given by suitable generalizations of the Jones knot
polynomial for the case of intersecting knots. These are the first
physical states of Quantum Gravity for which an explicit form is known
both in the connection and loop representations. Implications of this
result are also discussed.

\end{abstract}

\section{\bf Introduction}

The introduction of a new set of canonical variables for the
Hamiltonian treatment of General Relativity by Ashtekar\cite{As} has
opened new possibilities of achieving a Dirac canonical quantization
of the gravitational field. The new variables cast the dynamics of
General Relativity in terms of a connection rather than in terms of a
metric, making the phase space appear as imbedded in that  of a
Yang Mills theory. If one quantizes the theory taking a
polarization in which wavefunctions are functions of the Ashtekar
Connection $\Psi[A]$, one obtains the so called connection
representation\cite{As,JaSm}.
Another representation\cite{RoSm}, similar to the ones previously
introduced for Yang Mills theories\cite{GaTr}, is the Loop
Representation. In it,
wavefunctions are functionals on loop space $\Psi[\gamma]$.

The classical canonical variables of Ashtekar are a set of triads or
frame fields on the three manifold $\Sigma$ of a foliation of space
time, and are usually denoted by $\tilde{E}^a_i(x)$, where $a$ is a
spatial index on $\Sigma$ and $i$ is a flat Euclidean index, which can
be thought of as an $SO(3)$ index (the tilde denotes a density weight).
Canonically conjugate to these variables
is an $SO(3)$ connection, obtained by pulling back to $\Sigma$ the
self dual part of the spin connection of spacetime. It is usually
denoted as $A_a^i(x)$ and again $a$ is a tensor index on $\Sigma$ and $i$
is an $SO(3)$ index. The constraints equations of General Relativity,
with a cosmological constant $\Lambda$ are:
\begin{eqnarray}
{\cal G}^i =&& D_a \tilde{E}^{ai}\\
{\cal C}_b =&& \tilde{E}^{ai} F_{ab}^i=0\\
{\cal H} =&& \epsilon_{ijk} \tilde{E}^{ai} \tilde{E}^{bj} F^k_{ab}
-{\textstyle {\Lambda \over 6}}
\ut{\eta}_{abc}\epsilon^{ijk}\tilde{E}^{ai}\tilde{E}^{bj}\tilde{E}^{ck}
\end{eqnarray}
where $F_{ab}^i$ is the curvature of $A^i_a$.

When quantizing the theory in the connection representation, the
wavefunction are holomorphic functionals of the connection $\Psi[A]$
so that the connection is a multiplicative operator ${\hat A}^{i}_a
\Psi[A] = A_a^i \Psi[A]$ and the triad is a functional derivative
${\hat E}^{ai} \Psi[A] = {\delta \over \delta A_a^i} \Psi[A]$. To
promote the constraints to quantum wave equations one is faced with a
regularization and a factor ordering problem.

Two main factor orderings have been considered in the literature,
which basically consist in ordering the functional derivatives all to
the left or all to the right. We will call these two choices I and II
respectively and the explicit expressions for the constraints are
shown in table I.  The first one was considered by Ashtekar\cite{As} and
the second one by Jacobson and Smolin\cite{JaSm}.

Factor ordering I has the following features:

\noindent a) The algebra of constraints formally closes. (This result is only
formal, the factor ordering probably cannot be properly addressed
without the introduction of a regularization\cite{TsWo,FrJa}).

\noindent b) The diffeomorphism constraint {\it fails} to generate
diffeomorphisms on the wavefunctions, at least formally.

\noindent c) There exists a solution to the Hamiltonian constraint with a
cosmological constant, which is
also diffeomorphism invariant, given by\cite{Ko,So}:
\begin{equation}
\Psi_{\Lambda}[A] = {\rm exp} (-{\textstyle{6 \over \Lambda}}
\int \tilde{\eta}^{abc} Tr[A_a
\partial_b A_c +{\textstyle 2 \over 3} A_a A_b A_c])
\end{equation}
That is, the exponential of the Chern Simons form is a solution to the
Hamiltonian constraint, which can be very easily checked using the
relation:
\begin{equation}
{\delta \over \delta A_a^i} \Psi_{\Lambda}[A] = {\textstyle {3 \over \Lambda} }
\tilde{\eta}^{abc} F_{bc}^i \Psi_{\Lambda}[A]
\end{equation}
Moreover this function is invariant under diffeomorphisms, but at
least formally it is not annihilated by the diffeomorphism constraint,
which in this factor ordering fails to generate diffeomorphisms on the
wavefunctions. It is also a nondegenerate solution in the sense of
\cite{BrPu}.

\noindent d) It is from this factor ordering that one can obtain the Loop
Representation via the introduction of the transform\cite{RoSm,GaTr}
\begin{equation}
\Psi[\gamma] = \int d A\quad Tr( Pexp \oint \dot \gamma^a A_a) \Psi[A]
\label{transform}
\end{equation}
Consider how an operator $\hat O_\gamma$ in the loop
representation is obtained as the transform of $\hat O_A$ in the connection
representation:
\begin{eqnarray}
\hat{O}_\gamma \Psi[\gamma] & \equiv & \int dA \; Tr( Pexp
\oint \dot \gamma^a A_a) ) \hat O_A \Psi[A] \\
&=& \int dA \; (\hat O_A^\dagger Tr( Pexp
\oint \dot \gamma^a A_a) )  \Psi[A]
\end{eqnarray}
Assuming that $\hat O_A$ is self-adjoint with respect to the measure in the
transform, $\hat O_\gamma$ is the operator which acting on the kernel of the
transform has the same action as $\hat O_A$.
The loop representation \cite{RoSm,Ga} is based on factor ordering I and on the
assumption that $\hat A^i_a$ and $\hat{\tilde{E}}^{ai}$ are self-adjoint. But
when finding the transform of products, the factor ordering is reversed, i.e.
the constraints in the loop representation are based on the action of the
constraints on the kernel with factor ordering II. The opposite factor ordering
does not lead to the loop representation.

\noindent e) In particular, the proposed wavefunction can be transformed to the
loop representation, as was considered, for instance (for the
restricted case of nonintersecting loops) in\cite{Wi,CoGuMaMi,Sm}, its
transform being related to the Jones Polynomial. These wavefunctions
belong to the well known -and degenerate\cite{BrPu}- set of solutions to the
constraints of quantum gravity based on non-intersecting
loops\cite{JaSm,RoSm,Ga}

Factor ordering II has been considered elsewhere\cite{JaSm,BrPu}. In
this factor ordering it was found that the holonomy of the Ashtekar
connection (for the case of a smooth loop) was a solution to the
Hamiltonian constraint of the theory.  Moreover the diffeomorphism
constraint coincides with the generator of diffeomorphisms on the
wavefunctions.  However, this factor ordering is in principle endowed
with a difficulty: formally the algebra of constraints fails to close.
(This was already noticed in \cite{JaSm}).  To be more precise the
commutator of two Hamiltonians is "proportional" to a diffeomorphism,
as it should be, but due to the factor ordering the proportionality
factor appears {\it to the right}.  This means that if one has a
physical state $\Psi_{\rm ph}[A]$ (annihilated by both diffeomorphism and
hamiltonian
constraints) one could get a potential inconsistency in that the
equation:
\begin{equation}
[ \hat{\cal H}(M) , \hat{\cal H}(N)] \Psi_{\rm ph}[A] = 2 \hat{\cal C}_a (M
\partial_b
N - N\partial_b M) \hat{q}^{ab} \Psi_{\rm ph}[A]
\end{equation}
would have a left member identically zero and a nonvanishing right
member. This, plus other reasons lead us to consider in this paper factor
ordering I.

The main objective of this paper is to show that the following points
can be accomplished:

a) The construction of a regularized, factor ordered version of the
connection representation which is consistent in the sense expounded
above.

b) The construction of nondegenerate physical states in the connection
representation so developed.

c) An explicit calculation of the transform to the loop representation
of the solutions so constructed.

d) The discussion of up to what extent the loop and connection
representations are consistent with each other.

The organization of this article is as follows: in section 2 we will
show how a regularized version of the diffeomorphism constraint in
factor ordering I generates diffeomorphisms on the wavefunctions and
therefore $\Psi_{\Lambda}[A]$ becomes now a solution to all the constraints
of Quantum Gravity and therefore a physical state of the theory.  In
section 3 we introduce the generalization of the Jones Polynomials to
the case of intersecting loops. The idea is to compute the
transform of the physical state $\Psi_{\Lambda}[A]$
to the Loop Representation, including
the nontrivial case of intersecting loops. This is done in section 4.
The resulting polynomials, in spite of having an arbitrary number of
intersections, will automatically solve the complicated hamiltonian
constraint of Quantum Gravity in the loop representation.
In section 5 we discuss the issue of diffeomorphism invariance of the
solutions found and how the use of loops requires a regularization
that breaks diffeomorphism invariance.

\section{\bf Regulating the diffeomorphism constraint}

If we consider factor ordering I we have an almost satisfactory
situation at the formal level. The algebra of constraints closes at
the quantum level, there exists a solution to the Hamiltonian
constraint (which in addition is "nondegenerate" in the sense exposed
in \cite{BrPu}). However, a very unsatisfactory point arises when one
notices that (formally) the diffeomorphism constraint fails to
generate diffeomorphisms on the wavefunctions.  At this point it is
convenient to take into account the analysis of Tsamis and
Woodard\cite{TsWo} and Friedman and Jack\cite{FrJa} concerning the
need of a regularization for a consistent treatment of the factor
ordering problem and the constraint structure of Quantum Gravity.  We
will therefore regulate the diffeomorphism constraint in the same way
the hamiltonian constraint is usually\cite{JaSm,RoSm,Ga} regulated: by
point splitting. If we point split the diffeomorphism constraint:
\begin{equation}
\hat{\cal C}_{\epsilon}(\vec{N}) = \int d^3 x \int d^3 y N^a(x) f_\epsilon(x,y)
{\delta \over \delta A_b^i(x)} F_{ab}^i(y)
\end{equation}
where $f_\epsilon(x,y)$ is any even($f_\epsilon(x,y)=f_\epsilon(y,x)$)
regulator that in the limit $\epsilon \rightarrow 0$ goes to $\delta(x,y)$.
One can immediately check that,
\begin{equation}
\hat{\cal C}_{\epsilon}(\vec{N}) = \int d^3 x \int d^3 y N^a(x) f_\epsilon(x,y)
F_{ab}^i(y) {\delta \over \delta A_b^i(x)}
\end{equation}
since due to the symmetry of the regulator the extra term:
\begin{equation}
\int d^3x N^a(x) \int d^3y f_\epsilon(x,y) \partial_{[a} \delta(x,y)
\delta^b_{b]} \delta^i_i
\end{equation}
vanishes upon integration by parts.

That is, in the regularized version, the diffeomorphism constraint
generates diffeomorphisms in the factor ordering prescribed.
This simple calculation shows that factor ordering I is  consistent
for quantum gravity. It respects the symmetry of the
theory under diffeomorphisms, gives a correct closure (at least
formally) to the algebra of constraints and allows the construction of
a simple and nondegenerate physical state. This state is given by
expression (4), which is now annihilated by all the regularized
constraints of Quantum Gravity with cosmological constant.

It should be noticed that the solutions to the Hamiltonian constraint
known in the connection representation up to present
\cite{JaSm,Hu,BrPu} correspond to the factor ordering II. Therefore
they cannot be used as a starting point to construct solutions to the
Hamiltonian constraint of Quantum Gravity in loop space.

\section{\bf Knot polynomials for intersecting loops}

It is well known that the transform of the state (4) into the loop
representation is given by a knot polynomial closely related to the
Jones Polynomial. This calculation has been performed by several
different techniques.Witten\cite{Wi} considered a nonpertubative
approach whereas Smolin\cite{Sm} and
Cotta-Ramusino et al\cite{CoGuMaMi} checked perturbatively
Witten's claim (this latter method is the one to be considered
later on in this paper). In all these calculations however,
only smooth nonintersecting loops have been considered.

In Quantum Gravity however, one needs more generality. The Hamiltonian
constraint of Quantum Gravity in the Loop Representation trivially
annihilates all states with support only on smooth nonintersecting loops.
Unfortunately, since states with support on smooth nonintersecting
loops are also annihilated by the determinant of the spatial metric,
they become states of the gravitational field for an arbitrary value
of the cosmological constant (to see this notice that the only
difference between the Hamiltonian formalism with and without a
cosmological constant is a term proportional to the determinant of the
three metric). Therefore all these states are physical in the sense
that they are annihilated by all the constraints, but basically
correspond to spatially degenerate metrics \cite{BrPu}. Moreover, the
determinant of the three metric appears not only in the cosmological
constant term but in matter couplings in general\cite{AsRoTa}.

One can, however, perform the loop transform of the state (4) for the
case of loops with intersections. This will be one of the main points
of this paper. The result will be knot polynomials (appropriately
generalized to include intersections). These polynomials, in spite of
including an arbitrary number of arbitrary-order intersections, will
however, manage to solve the complicated hamiltonian constraint of
quantum gravity. The reason for this is that we will obtain them as
loop transforms of a functional that solves the hamiltonian
constraint in the connection representation!

We are therefore interested in obtaining the loop transform allowing
the loops to have intersections. For this we will first need to
generalize notions of knot polynomials to the intersecting case. This
has received some attention in the past\cite{KaEn,Ga92}. We now
briefly sketch how to generate knot polynomials with
intersections from notions of the Braid Group.

A standard technique for constructing knot polynomials is to start
from the Braid Group. The Braid Algebra $B_n$ is composed of elements
$g_i$, with $ 0<i<n$ (as depicted in figure 1) that satisfy:
\begin{eqnarray}
g_i \; g_j =&& g_j \; g_i\qquad {\rm for}\;\; |i-j|>1\\
g_i \; g_{i+1} \; g_i =&& g_{i+1} \; g_i \; g_{i+1}
\end{eqnarray}
Each element of $B_n$ represents a braid diagram composed by lines,
called strands lying on a plane and moving each around the other. If
$g_i$ represents an over crossing of the line $i$ and $i+1$ the
corresponding undercrossing is represented by $g_i^{-1}$. Two braids
are equivalent if they may be transformed into each other by smooth
deformations of the strands in $R^3$, leaving their endpoints fixed.
To proceed from braids to knots one identifies the top and bottom ends
of the braid. Two knots are equivalent if they differ by a finite
sequence of moves known as Markov moves\cite{Bi}.
The Braid Algebra can be enlarged to consider the case of Braids with
intersections\cite{KaEn,Sm}. One just introduces a new generator $a_i$
representing a 4-valent rigid vertex . If one wishes to consider
intersections of more than two braids at each point the algebra has to
be enlarged further\cite{Ga92}.

The extended braids are also subject to relations following from
equivalence under smooth deformations in $R^3$. They are:

\begin{eqnarray}
a_i g_i =&& g_i a_i\\
g_i^{-1} \; a_{i+1} \; g_i =&& g_{i+1} \; a_i \; g_{i+1}^{-1}
\end{eqnarray}
and:
\begin{eqnarray}
[ g_i, a_j]=0 \qquad [a_i, a_j]=0 \qquad|i-j|>1
\end{eqnarray}

One can find matrix representations for the Braid
Algebras\cite{Jo,Tu,Gu}.  From these representations one can derive
skein relations for the knot polynomials. The one that yields the
Jones Polynomial (in the nonintersecting case) is related with
associating to each generator $g_i$ a $2^n \times 2^n$ matrix acting
on the linear space $V(n)=V_1 \otimes V_2 \otimes...\otimes V_n$ where
$V_i$ is a two dimensional space corresponding to the strand $i$. More
precisely $G_i$ will be represented by:
\begin{equation}
G_i = q^{1/4} (I \otimes....\otimes K \otimes...\otimes I)
\end{equation}
where $q$ is an arbitrary complex number, $I$ is the $2 \times 2$
identity matrix and the matrix $K$, which acts on $V_i \otimes V_{i+1}$,
is given by:
\begin{eqnarray}
K \: &=&
\: \left(
\begin{array}{cccc}
1 & 0 & 0 & 0\\
0 & \; 1-q^{-1} \; & \;q^{-1/2}\; & 0\\
0 & q^{-1/2} & 0 & 0 \\
0 & 0 & 0 & 1
\end{array} \right)
\end{eqnarray}
An extended representation including the vertex generators is given by
the $2^n \times 2^n$ matrices \cite{Ga92}:
\begin{equation}
A_i = I \otimes ...\otimes A \otimes ... \otimes I
\end{equation}
where $A$ is given by the matrix acting on $V_i \otimes V_{i+1}$:
\begin{eqnarray}
A \: &=&
\: \left( \:
\begin{array}{cccc}
1 & 0 & 0 & 0\\
0 & a & \;(1-a) q^{1/2} \;& 0\\
0 & \; (1-a) q^{1/2} \; & \; 1 - (1-a) q\; & 0\\
0 & 0 & 0 & 1
\end{array} \; \right)
\end{eqnarray}
where $a$ is another complex  parameter.

The skein relations result from the following identities satisfied by
the  $G_i$ and $A_i$ matrices in this representation:
\begin{eqnarray}
q^{1/4} G_i &-& q^{-1/4} G_i^{-1} - (q^{1/2} -q^{-1/2}) I_i = 0\\
A_i &=& q^{1/4} (1-a) G_i^{-1} + a I_i
\end{eqnarray}

To construct a regular isotopy invariant link polynomial one defines
the enhancement matrix:
\begin{equation}
M_n = \mu_1 \otimes \mu_2 \otimes ... \otimes \mu_n
\end{equation}
with:
\begin{eqnarray}
\mu_i &=& \;
\left(
\begin{array}{cc}
\;q^{-1/2} \; & 0 \\
0 & \; q^{1/2} \;
\end{array}
\right)
\end{eqnarray}
Then one can show that:
\begin{equation}
F(B) = Tr[B M_n]
\end{equation}
is a regular isotopy link invariant, where the trace is taken in the
vector space $V(n)$,  and $B$ is a matrix representing an arbitrary
element of $B_n$.

Given the diagrams of figure 2, we can now write the skein relations
satisfied by $F(q,a)$ as:
\begin{eqnarray}
F_{\hat{L}_+} &=& q^{3/4} F_{\hat{L}_0}\\
F_{\hat{L}_-} &=& q^{-3/4} F_{\hat{L}_0}\\
q^{1/4} F_{L_+} -q^{-1/4}  F_{L_-} &=& (q^{1/2} -q^{-1/2}) F_{L_0}
\label{skein3} \\
F_{L_I} &=& q^{1/4} (1-a) F_{L_-} + a F_{L_0}\\
F_{0} &=& 1
\end{eqnarray}
the last being the standard normalization condition on the unknot.

This ends our discussion of the generalized Jones polynomial for
intersecting loops. It is clear that for non intersecting loops it
reduces to the standard form of the Jones Polynomial if one multiplies
our results by $q^{-3/4 w(L)}$ where $w(L)$ is the writhing of $L$.
The reason for this difference is that the Jones Polynomial is ambient
isotopic invariant whereas we are more interested in polynomials that
are regular isotopic invariant for reasons we will see in the next
section.

\section{\bf Skein relations for the physical state}

Let us consider the loop transform of the physical state
$\Psi_{\Lambda}[A]$. For the nonintersecting case similar calculations were
performed by Smolin\cite{Sm} and Cotta-Ramusino et al.\cite{CoGuMaMi}.
The first calculation was actually performed by Witten\cite{Wi} but
with different techniques.

The loop transform is given by:
\begin{equation}
\Psi[\gamma]= \int dA\quad Tr[U(\gamma)] {\rm exp}(-{\textstyle {12\over
\Lambda}}
S_{CS}[A])
\end{equation}
where $Tr[U(\gamma)]=Tr[ P {\rm exp} (\oint A_a \dot \gamma^a)]$ and
$S_{CS}[A]= \frac{1}{2} \int \tilde{\eta}^{abc} Tr[A_a\partial_bA_c
+ \frac{2}{3}A_aA_bA_c]$
is the Chern-Simons action.

We now consider the variation of this expression when a small loop of
area $\Sigma^{ab}$ is appended to the loop $\gamma$. Let us first
consider the case without intersections. We get:
\begin{equation}
\Sigma^{ab} \Delta_{ab}(x) \Psi[\gamma] = \int dA \quad \Sigma^{ab}
F_{ab}^i(x) Tr[\tau^i U(\gamma_x^x)] exp(-{\textstyle {12 \over
\Lambda}} S_{CS})
\end{equation}
where $\Delta_{ab}$ is the area derivative\cite{GaTrAd,BrPu92}, $\tau^i$ is
one of the generators of $SU(2)$ and we have used $\Delta_{ab}(x)
Tr[U(\gamma)] = F_{ab}^i(x) Tr[\tau^i U(\gamma_x^x)]$ and $\gamma_x^x$
is the loop with origin at the point $x$.

Using the relation (5) and integrating by parts, one obtains:
\begin{equation}
{\textstyle -{\Lambda \over 6}}
\int dA \quad \Sigma^{ab} \eta_{abc} \int dy^c \delta(x-y)
Tr[ \tau^i U(\gamma_x^y) \tau^i U(\gamma_y^x)] exp(-{\textstyle {12 \over
\Lambda}} S_{CS})
\end{equation}

The integral depends on the volume factor
\begin{equation}
\Sigma^{ab} \eta_{abc} dy^c \delta(x-y)
\end{equation}
which depending on the relative
orientation of the two-surface $\Sigma^{ab}$ and the differential
$dy^c$ (which is tangent to $\gamma$), can lead to  $\pm 1$ or zero.
(This expression should really be regularized. We have absorbed
appropriate extra factors in the definition of the cosmological
constant so to normalize the volume to $\pm 1$).
Consequently, depending on the value of the volume there are  three
possibilities:
\begin{eqnarray}
\delta \Psi[\gamma] &=&0\\
\delta \Psi[\gamma] &=& \pm {\textstyle {\Lambda \over 8}} \Psi[\gamma]
\end{eqnarray}
These equations can be diagrammatically interpreted in the following
way:
\begin{equation}
\Psi[\hat{L}_\pm]-\Psi[\hat{L}_0] = \pm {\textstyle {\Lambda \over 8}}
\Psi[\hat{L}_0]
\end{equation}
When the volume element vanishes it corresponds to a variation that
does not change the topology of the crossing.

Let us now consider the case of a point where there is an
intersection. As before, we
consider an infinitesimal deformation of the loop consisting in the
addition of a small closed loop, in this case at the point of
intersection (see figure 3):
\begin{equation}
\Sigma^{ab} \Delta_{ab}(y) \Psi[\gamma] = {\textstyle {\Lambda \over 6} }
\int dA \Sigma^{ab}
\eta_{dab} Tr[\tau^i U_{23}(\gamma_y^y)
U_{41}(\gamma_y^y)] {\delta \over \delta
A_d^i(y)} {\rm exp}({\textstyle -{12 \over \Lambda}} S_{CS})
\end{equation}
Again integrating by parts and
choosing the element of area $\Sigma^{ab}$ parallel to segment 1-2 so that the
contribution of the functional derivative corresponding to the action on the
segment 1-2 vanishes (since the volume element is zero) we get:
\begin{eqnarray}
\Sigma^{ab} \Delta_{ab} \Psi[\gamma] &=&
-{\textstyle {\Lambda \over 6}} \int dA
\Sigma^{ab} \eta_{abc} \int dv^c \delta(y-v) \times \nonumber\\
&&\times Tr[ \tau^i
U_{23}(\gamma_y^y) \tau^i U_{41}(\gamma_y^y) ] {\rm exp}
({\textstyle -{12 \over \Lambda}} S_{CS})
\end{eqnarray}
Making use of the Fierz identity for $SU(2)$:
\begin{equation}
\tau^i{}^A_B \tau^i{}^C_D = -{1 \over 2} \delta^A_D \delta^C_B +{1\over 4}
\delta_B^A \delta_D^C
\end{equation}
one finally gets:
\begin{eqnarray}
&&\Sigma^{ab} \Delta_{ab} \Psi[\gamma] = \\
&=& {\textstyle {\Lambda \over 12}} \int dA \;
\Sigma^{ab} \eta_{abc} \int dv^c \delta(y-v) Tr[U_{23}(\gamma_y^y)]
Tr[U_{41}(\gamma_y^y)] {\rm exp}({\textstyle -{12 \over \Lambda}}
S_{CS}) \nonumber\\
&-& {\textstyle {\Lambda \over 24 } }
\int dA \; \Sigma^{ab} \eta_{abc} \int dv^c \delta(y-v)
Tr[U_{23}(\gamma_y^y) U_{41}(\gamma_y^y)]
{\rm exp}( {\textstyle -{12 \over \Lambda}} S_{CS})\nonumber
\end{eqnarray}
where we have called $U_{ij}(\gamma_{x_1}^{x_2})$ the holonomy from point $x_1$
to $x_2$ traversing through lines $i$ and $j$.

These relations can be interpreted as the following skein relation
for the intersection.
\begin{eqnarray}
\Psi[L_\pm] &=& (1 \mp {\textstyle {\Lambda \over 24}})
\Psi[L_I] \pm {\textstyle {\Lambda \over 12}} \Psi[L_0]\label{plusminus}
\label{rel}\\
\Psi[\hat{L}_\pm] &=& (1 \pm {\textstyle {\Lambda \over 8}}) \Psi[\hat{L}_0]
\end{eqnarray}
In order to compare with the link polynomials we must first notice
that the results we have obtained correspond to a linear
approximation, since we have only considered an infinitesimal
deformation of the link. In order to consider a finite deformation we
would have to consider higher order derivatives of the wavefunction.

It is convenient to rewrite the relations obtained in such a way that
the correspondence with those of the Jones Polynomials
in the intersecting case is manifest. To do this we notice that the factor
$(1+{\Lambda \over 8})$ plays the role of $q^{3/4}$ and therefore in the
linearized case if we define $q$ as $q=e^k$, then $k={\Lambda \over
6}$. Inverting the relation (\ref{rel}) we get:
\begin{equation}
\Psi[L_I]=(1\pm {\textstyle {\Lambda \over 24}}) \Psi[L_\pm] \mp {\textstyle
{\Lambda \over 12}}\Psi[L_0] \label{reli}
\end{equation}
which allows us to recognize that the value of the variable $a$ of the
Generalized Jones Polynomial is $a=1 - e^{-{\Lambda \over 12}}$ which to
first order yields $a ={\Lambda \over 12}$.

Expression (\ref{skein3}) relating $\Psi[L_+]$ and $\Psi[L_-]$ can be
obtained in this case by combining eqs. (\ref{reli}). Previous
derivations of these expressions (for the nonintersecting case)
\cite{CoGuMaMi}are somewhat misleading since the functional derivative
in the case where there is no intersection only can act on one side of
the crossing and therefore gives a vanishing contribution.

So we see that the Generalized Jones Polynomials, introduced in the last
section for loops with double self-intersections from the Braid Group,
are actually the loop transforms of a physical nondegenerate quantum
state of the gravitational field defined by values of $q$ and $a$
that, to first order in perturbation theory coincide with the ones
presented above.

In this sense we can therefore say that they are annihilated by the
constraints of quantum gravity in the loop representation and are
therefore physical states of the gravitational field. The issue of
finding physical states of the gravitational field that were
nondegenerate received a lot of attention recently and great effort
went into the construction of examples of such states
\cite{JaSm,Hu,BrPu,BrGaPu}. Since we started from a
$\Psi_{\Lambda}[A]$ which was nondegenerate in the connection
representation, its transform to loop space will also be
nondegenerate. In this paper we have considered a restricted transform
in the subspace of loops with double intersections. A simple extension
of the calculations presented in this paper, by including higher order
intersecting loops in the transform would allow the construction of a
physical state in loop space that is not annihilated by the
determinant of the three metric and which would therefore be
nondegenerate.

\section{\bf Discussion and conclusions:}

We have found a solution to all the constraints of Quantum Gravity in
both the connection and the loop representations. This allows for the
first time to analyze the consistency between both representations,
and to some extent clarify the properties of the loop transform. The
existence of a transform has received great attention recently due to
the attempts to formulate a rigorous definition for it in several
theories by Ashtekar and Isham\cite{AsIs}.

In connection with this point a surprising fact arises, which was
already known to people working in Chern-Simons theories\cite{Wi}:
although the state $\Psi_{\Lambda}[A]$ in the connection representation is
diffeomorphism invariant (and annihilated by the diffeomorphism
constraint), the transformed state $\Psi_\Lambda[\gamma]$ is only a {\it
regular} isotopic invariant. In fact the physical
state can be written as:
\begin{equation}
\Psi_\Lambda[\gamma] = (1-{\textstyle{\Lambda \over 8}})^{w(\gamma)} P(\gamma)
\end{equation}
where $w(\gamma)$ is the writhing of $\gamma$ and $P(\gamma)$ is the
Jones Polynomial. The Jones Polynomial {\it is} diffeomorphism
invariant (it is ambient isotopic invariant in the knot theory
language), however the writhing number is not. Therefore we see that
the failure to obtain a diffeomorphism invariant solution is
concentrated in the first factor. One could try to fix this problem by
eliminating this first factor and directly try to propose the Jones
Polynomial as a solution to the constraints. This however, fails. The
Hamiltonian constraint has a nontrivial action on the writhing number.
What is a solution of all the constraints is $\Psi_\Lambda[\gamma]$ and not
$P(\gamma)$.

A reasonable question therefore is to ask what happened to the
consistency of both representations. One started from a state that was
diffeomorphism invariant and annihilated by the diffeomorphism
constraint and by transforming to the loop representation
diffeomorphism invariance was lost. This occurs since the writhing
depends on how the three dimensional knot is projected into a plane.
For a fixed projection, a diffeomorphism of the knot in three
dimensions can change the writhing since $\hat{L}_\pm$ and $\hat{L}_0$
contribute differently. The ambiguity is not present if one considers
bands instead of loops (see figure 4). A prescription which given a
loop produces a band is called framing. Therefore in order to obtain a
diffeomorphism invariant state one needs a diffeomorphism invariant
framing. Evidently, the choice of framing was hidden in the implicit
measure $dA$ in the space of connections used in the transform and
this is how diffeomorphism invariance was lost.

At this point it is clear that the very use of loops is generating
problems with invariance under diffeomorphisms.  This can be seen with
greater clarity if one considers a simpler example, the case of an
Abelian theory. If one considers the loop transform of an Abelian
Chern-Simons functional:
\begin{equation}
\int dA\; {\rm exp}(k \int d^3x   \tilde{\eta}^{abc} A_a \partial_b
A_c) \times {\rm exp}( i \oint
\dot{\gamma}^a A_a )
\end{equation}
the integral can be explicitly computed since it is a gaussian and
corresponds to the exponential of the Gauss Self Linking number:
\begin{equation}
\oint ds \oint dt \dot{\gamma}^a(s) \dot{\gamma}^b(t) \epsilon_{abc}
{(\gamma(s) -\gamma(t))^c \over |\gamma(s) -\gamma(t)|^3}
\end{equation}
This quantity is ill defined when $\gamma(s)=\gamma(t)$. Therefore it
has to be regularized (These issues have extensively been discussed in
\cite{Gu} and references therein). This is again accomplished by means
of a framing and leads to a functional of bands rather than of knots.
This problem evidently comes from the singular (distributional)
character of loops as functions of the three manifold and the
associated need of a regularization.

We would like to propose a possible solution to this problem based on
the replacement of the loops by smooth nonsingular objects. We will
only illustrate the point with a brief discussion of the abelian
case, the nonabelian generalization and a detailed discussion of the
point in general being beyond the scope of this article.  The key
point is to notice that the only way "loops" have entered the
formalism is via the holonomy, in which they are represented by
divergence-free vector densities $\tilde{X}^a(x)$, defined by
$\tilde{X}^a(x) =\oint ds
\dot{\gamma}^a(s) \delta^3(x-\gamma(s))$ which we will call ``loop
coordinates'' (they have also been called ``form
factors''\cite{AsRoSm}).  In terms of them, the holonomy is a three
dimensional integral $ h[\gamma,A]= \int d^3x A_a(x)
\tilde{X}^a(x) $. Now one could consider a representation based,
instead of on loops, {\it  on
smooth divergence free vector densities} $\tilde{X}^a(x)$,
which is inherently free of the mentioned singularities. For
instance, the transform to this representation of the Chern-Simons
wavefunction would be the exponential of:
\begin{equation}
\int d^3x \int d^3y \tilde{X}^a(x) X_a(y)
\end{equation}
where $X_a(y)$ is the ``potential'' defined by $\partial_{[b}
X_{a]}(y) =\eta_{bac} \tilde{X}^c(y)$ (whose existence is guaranteed
due to the fact that the $\tilde{X}^a$ are divergence free; one can
also check that if one considers the $\tilde{X}^a(x)$ as given by the
expression in terms of the loops this expression yields the linking
number). This expression is completely nonsingular, well defined and
diffeomorphism invariant.

This strongly suggests that a nonabelian generalization of this
construction, based on coordinates on loop space formed by multivector
densities\cite{GaLe,DiGaGrLe,BrGaPu} could analogously solve the problem
of the loss of diffeomorphism invariance of the loop transformed
expressions of the physical states here introduced.

A natural calculation that arises from the issues discussed here is to
attempt to show in an explicit way if the wavefunctions presented here
in the loop representation are annihilated by the Hamiltonian
constraint in that representation directly. The calculation is long
and exceeds the scope of this paper. However one can immediately draw
some qualitative conclusions from it by looking at the first terms in
the expansion. They correspond to known knot invariants\cite{GuMaMi},
on which one
can explicitly compute the action of the Hamiltonian constraint in the
loop representation. This allows, by requiring consistency at each
order, to draw conclusions about the behaviour of each knot invariant
under the action of the hamiltonian constraint. The results of this
analysis will be presented elsewhere. In particular, to zeroth order
in the cosmological constant, the polynomial becomes a function of the
number of connected parts of the link. This term is immediately a
solution of the Hamiltonian constraint (in vacuum) since it is
annihilated by the area derivative. It can also be checked that it is
not annihilated by the determinant of the three metric (for instance
for three loops, applying formula (12) of reference \cite{BrGaPu}),
and therefore it is the simplest example of a nondegenerate physical
state of the gravitational field.

In a forthcoming paper we will analyze how, by looking order by order
in the expansion of $\Psi_\Lambda[\gamma]$ we can actually find an
infinite set of solutions to all the constraints of quantum gravity
(with vanishing cosmological constant) in the loop representation,
including as a particular case a remarkably simplified derivation of
the results of reference \cite{BrGaPu}.

\section{\bf Acknowledgements}
\nonumber

We wish to especially thank Abhay Ashtekar and Lee Smolin for many
fruitful discussions. R.G. also thanks Abhay Ashtekar, Lee Smolin,
Karel Kucha\v{r} and Richard Price for hospitality and financial
support during his visit to Syracuse University and The University of
Utah. This work was supported in part by grant NSF PHY 89 07939 and
NSF PHY 90 16733 and by research funds provided by the Universities of
Syracuse and Utah. Financial support was also provided by CONICYT,
Uruguay.

\mediumtext
\begin{table}
\setdec 0.000
\caption{Two factor orderings for the constraints of
Quantum Gravity in terms of
Ashtekar's variables}
\begin{tabular}{cc}
Factor Ordering I& Factor Ordering II\\
\tableline
$\hat{\cal G}^i =D_a {\delta\over\delta A_a^i}$&
$\hat{\cal G}^i=D_a {\delta\over\delta A_a^i}$\\
$\hat{\cal C}_b = {\delta \over \delta A_a^i} F_{ab}^i$&
$\hat{\cal C}_b = F_{ab}^i {\delta \over \delta A_a^i}$\\
$\hat{\cal H} = \epsilon_{ijk} {\delta \over \delta A_a^i}
{\delta \over \delta A_b^j} F_{ab}^k -{\Lambda \over 6}
\epsilon_{ijk} \ut{\eta}^{abc}
{\delta \over \delta A_a^i}
{\delta \over \delta A_b^j}
{\delta \over \delta A_c^k}$&
$\hat{\cal H} = \epsilon_{ijk} F_{ab}^k {\delta \over \delta A_a^i}
{\delta \over \delta A_b^j} -{\Lambda \over 6}
\epsilon_{ijk} \ut{\eta}^{abc}
{\delta \over \delta A_a^i}
{\delta \over \delta A_b^j}
{\delta \over \delta A_c^k}$\\
\end{tabular}
\end{table}

\figure{Graphic picture of the Braid Group relations (12-16) for the
case of braids without and with intersections.}

\figure{Knot configurations involved the skein relations (43-45).}

\figure{Addition of a small closed loop for the calculation of the
skein relations in the intersecting case.}

\figure{A depiction of regular isotopic invariance. Both bands can be
associated to loops that are related by a diffeomorphism. However the
corresponding bands are not diffeomorphically related.}

\end{document}